# Large phase-transition-induced magnetic anisotropy change in (Co/Pt)$_2$/VO$_2$ heterostructure

Guodong Wei,[1,‡] Xiaoyang Lin,[1,2,3,‡,*] Zhizhong Si,[1] Yanxue Chen,[4] Sylvain Eimer,[5] and Weisheng Zhao[1,2,3,*]

[1] Fert Beijing Research Institute, School of Microelectronics & Beijing Advanced Innovation Center for Big Data and Brain Computing (BDBC), Beihang University, Beijing 100191, China

[2] Beihang-Goertek Joint Microelectronics Institute, Qingdao Research Institute, Beihang University, Qingdao 266000, China

[3] Hefei Innovation Research Institute, Beihang University, Xinzhan Hi-tech District, Anhui 230013, China

[4] School of Physics and State Key Laboratory of Crystal Materials, Shandong University, Jinan 250100, China

[5] Centre de Nanosciences et de Nanotechnologies, University of Paris-Sud, Université Paris-Saclay, Orsay 91405, France

‡These two authors contributed equally.

* Corresponding Authors: XYLin@buaa.edu.cn (X.Y.L), weisheng.zhao@buaa.edu.cn (W.S.Z)

**Abstract:**

We report the phase-transition controlled magnetic anisotropy modulation in the (Co/Pt)$_2$/VO$_2$ heterostructure, where VO$_2$ is introduced into the system to applied an interfacial strain by its metal-insulator transition. A large reversible modulation of the perpendicular magnetic anisotropy (PMA) reaching 38 kJ/m$^3$ is observed during this process. The calculated energy density variation of interfacial anisotropy reaches 100 mJ/m$^2$, which shows significant advantage over traditional modulation strategies. Further experimental results including magnetization change versus temperature, strain buffered modulation and pre-strained sample comparison prove that the interfacial coupling between VO$_2$ and PMA layers plays a crucial role in this modulation. This work, demonstrating the great potential of phase-transition material in efficient magnetic anisotropy modulation, would benefit the exploration for low-power consumption devices.

In recent years, researchers have devoted great efforts on the development of spintronic devices based on tunnel magnetoresistance (TMR)[1–3], spin-transfer torque (STT)[4,5], spin-orbit torque (SOT)[6–8] and other novel effects[9,10] to confront the advent of the post-Moore era. Newly developed devices have shown a growing trend to rely on the PMA technology for considerations of thermal stability and device minimalization[11]. The Co/Pt multilayer is one of the most widely studied PMA systems owing to its advantages in strong anisotropy and easy preparation[12–15]. In the process of realizing its functionalities, one major concern is to find an energy-efficient way to manipulate the magnetic properties of the sysytem[9]. Therefore, strain[16,17] and charge[18] based voltage control strategies have been widely studied over the past decade. However, it is hard to achieve a pronounced regulation of PMA by relatively small voltage in conventional multiferroic systems, either restricted by the dielectric constant of the insulating layers or limited by the substrate clamping effect for thin films. As a result, new modulation strategies are still highly desired.

Like ferroelectric material, phase-transition materials can also induce interfacial strain by the crystal structure change when contacted with other materials, thus a strain-mediated modulation can be expected when it is combined with spintronic materials[19]. What is more, as the phase-transition processes are usually very intense around the critical point, it is hopeful to achieve a strong regulation effect only using a small energy to trigger it. In addition, because the lattice type is totally changed in this process, the interfacial-strain could be much stronger than that induced by crystal structure distortion. As a typical representation, $VO_2$ exhibits a complex property change when it transforms from a monoclinic (M1) insulator into a rutile (R) metal at a critical temperature around 340 K (Fig. 1a)[20]. The phase-transition can also be induced by various kinds of methods like strain, electrical gating and photonic inspiring[20–22]. It is noteworthy that the inherent phase-transition time scale of $VO_2$ reaches 100 fs order[23], which means a VO$_2$ phase-transition based modulation of PMA also has a great potential for ultra-fast spintronic devices[24,25].

In this article, we report the large PMA modulation achieved in the (Co/Pt)$_2$/VO$_2$ heterostructure. The phase-transition of VO$_2$ is utilized to induce an interfacial-strain applied to PMA layers. Anomalous Hall effect (AHE) and magnetic hysteresis loops have been used to illustrate the reversible PMA variation during this process. The energy density change of perpendicular anisotropy and interfacial anisotropy reaches 38 kJ/m$^3$ and 100 mJ/m$^2$, respectively. Further experimental results including magnetization variation versus temperature, strain buffered modulation and pre-strained sample comparison confirm that the interface coupling answers for this modulation. This finding opens an interesting prospect for exploiting higher efficiency control of PMA by phase-transition for the next generation spintronic devices.

Fig.1 (a) shows the schematic drawing of the phase-transition controlled modulation of magnetic anisotropy (PCMA) heterostructure sample used in this work. The epitaxial VO$_2$ thin films were deposited on TiO$_2$ (100) substrates by a pulsed laser deposition (PLD) technique under a base pressure of 2.0 Pa at 500℃. The laser fluence and repetition rate were fixed at 4 J/cm$^2$ and 2 Hz, respectively. The VO$_2$ thickness was set to be 20 nm to get a proper modulation effect. We chose TiO$_2$ (100) as the substrate since the mismatch between the lattice constant of VO$_2$ (b=4.554 Å, c=2.856 Å) and that of TiO$_2$ (b=4.593 Å, c=2.959 Å) is less than 3.5% at grow temperature, so an epitaxial growth can be expected. Furthermore, our previous work suggests that better modulation effect can be achieved on TiO$_2$ (100) oriented substrates, owing to its larger lattice change as compared with other low index surfaces during the phase-transition. The morphology of the film at every stage was measured by atomic force microscope (AFM). As shown in Fig.1 (b), the sample surface after VO$_2$ deposition is relatively flat

(surface average roughness lower than 0.3 nm) which guarantees the successful growth of the PMA layers. To characterize the phase-transition properties of $VO_2$, the temperature dependence of the film resistance was investigated for $VO_2$ using the two-probe method. As shown in Fig.1 (c), electrical measurements were performed from 300 to 360K. A typical hysteresis cycle can be observed, showing the transition temperature upon heating is around 340K. The PMA layers were deposited by magnetron-sputtering at room temperature, which is made up of double Pt/Co unites with 1.8 nm thick Pt layer and 0.6 nm thick Co layer. The heterostructure were then covered with 5nm of Ta as a protective layer to prevent the oxidation of the films. After the entire film structure was prepared, the surface showed a polycrystalline morphology as shown in Fig.1 (c). The distribution width of the grain size is around 10nm forming a relatively smooth surface (surface average roughness lower than 1 nm) which meets the demand for magnetic recording.

The Hall resistance ($R_{Hall}$) was measured to investigate the magneto-transport properties of the PCMA film. Fig.2 shows the $R_{Hall}$-H curves of the sample at 300 and 360K. Obvious difference can be observed between the results before and after the phase-transition of $VO_2$ happens. It is found that the $R_{Hall}$-H curve of the sample shows hysteresis and nonlinear slop at 300K. The square shape of the Anomalous Hall effect signal indicates that the film has a perpendicular magnetic easy axis. While, a rounded $R_{Hall}$-H curve without hysteresis is observed at 360K after the phase-transition happens. The results show that the phase change of $VO_2$ can indeed regulate the PMA properties of the sample. Furthermore, this modulation is totally reversible as the squareness is recovered after the temperature decreased back to 300K.

To further investigate the magnetic modulation effect in this heterostructure. Magnetization measurements were then performed by a vibrating sample magnetometer (VSM) to measure the saturation magnetization ($M_s$) and obtain other magnetic information. The value of $M_s$ is determined to be 2030 emu/cm³ at 300K, which is greater than the value of pure Co film. This enhancement in the Co/Pt multilayers is reported owing to the polarization of Pt at the Co/Pt interface[26]. The magnetization of the heterostructure was measured as a function of temperature to investigate the modulation effect. As illustrated in Fig. 3 (a), the magnetization decreases with the temperature increase, owing to the $M_S$ out-of-plane (OOP) becomes larger in this process (see inset OOP). It is noteworthy that the decrease tendency is quite different before and after 340K which is consistent with the critical temperature of $VO_2$ phase-transition (Fig.1 (d)). This phenomenon indicates that the magnetism variation is indeed related to the phase-transition of the $VO_2$ layer. The insets of Fig. 3(a) give the hysteresis loops of the film measured with small magnetic field applied. The OOP results are consistent with the data obtained in the magneto-transport measurements. While no difference can be found between the results measured at 300 and 360K in-plane (IP). We also change the angles between the magnetic field and the sample, and the outcomes are always the same. Although the inner mechanism here needs further exploration, the results show that this phase-transition induced modulation could effectively change the magnetic property out-of-plane while maintain the magnetic properties in plane.

To evaluate the modulation effect on magnetic anisotropy in the PCMA sample, the magnetic parameters of the heterostructure before and after the phase-transition are processed and summarized in Table 1. We analyzed the IP and OOP hysteresis curves of the film and obtained the effective perpendicular anisotropy energy density, $K_{eff}$, which decreases from 76 kJ/m³ to 38 kJ/m³. The ratio of the magnetic anisotropy energy change[27], defined as $\Delta K_{eff}/(2K_{eff, ave})=(K_{eff, 300K}- K_{eff, 360K})/(K_{eff, 300K}+ K_{eff, 360K})$, reaches 33%, which is comparable to the reported magnetic anisotropy changes induced by electric field[28–31]. To separate the bulk and interfacial contributions of the anisotropy, $K_{eff}$ can be written as

$K_{eff}=K_b-2\pi M_S^2+K_i/t$ [27], where $K_b$ is the bulk crystalline anisotropy, $K_i$ is the interfacial anisotropy, $-2\pi M_s^2$ is the demagnetizing field and $t$ is the thickness of magnetic film. Because the change of saturation magnetization is quite small (around 1%), the corresponding modulation of demagnetization is very weak (around 3%). Considering the contribution from $K_b$ is neglectable in Co/Pt thin films[26], the significant modulation of perpendicular anisotropy energy is mainly dominated by the variation of the surface anisotropy energy. The energy density reaches 100 μJ/m$^2$, calculated as $K_i=(K_{eff}+2\pi M_s^2)t$, which needs an electric field at the order of 5*10$^7$ V/m to get a similar modulation effect in a voltage-controlled magnetic anisotropy (VCMA) system (Scaling 1800 fJ/Vm, see refs.[32–34]). Notice that the breakdown electric-field of normally used dielectric materials are around 10$^6$ V/m, which reflects the superior performance of the PCMA heterostructure.

As mentioned above, it is assumed that a magneto-elastic effect plays a crucial role in the origin of this modulation. The mechanism can be intuitively understood though a physical image of the interaction between the interfacial strain and the electron orbital hybridization. It has been demonstrated experimentally that Pt 5d-Co 3d hybridization, which is localized at the Co/Pt interface, causes the enhancement of the orbital magnetic moment, and leads to the PMA effect via spin-orbital coupling[35,36]. Theoretical calculation supports that the tensile and compressive stress along the z direction will lead to the electron occupancy change in the states between out-of-plane 3z$^2$-y$^2$ and in-plane x$^2$-y$^2$ orbitals[37]. Therefore, the phase-transition can directly modulate the PMA strength of the Co/Pt multilayers though an interfacial strain caused by the lattice change during the phase-transition of VO$_2$. Further discussion requires first principles calculations.

To confirm this strain mediated assumption, a reference sample was prepared with 3 nm Ta inserted into the film as a strain buffer layer between VO$_2$ and PMA layers. In this case, the distance between the upper layer of VO$_2$ and Co/Pt inter face is above 4.8 nm, i.e. around the limit of interficial strain effect. The *in situ* longitudinal magneto-optic Kerr effect (MOKE) measurement was used to investigate the magnetic property change at different temperatures. As illustrated in Fig. 4(a), only a faint shrinking of coercive field can be detected with the temperature increase, and the sloops keep their squareness during this process. However, for the ordinary sample without the insertion layer, a significant variation of saturation field ($H_S$) is observed as illustrated in Fig.4 (b). When the temperature is rised above 328K, the curves become linear without hysteresis like a superparamagnetic sloop. The sharp contrast between the samples with and without strain buffer layer confirms that the modulation effect in the PCMA sample relies on the interface coupling between VO$_2$ and PMA layers.

In this sense, the modulation effect can be further studied by a comparation between the ordinary sample and a pre-strained one. The pre-strained sample is prepared with the PMA layers deposited at 373 K, under which condition the VO$_2$ is at R state. Thus, after deposition, the sample will return to M1 state with a pre-strained effect. As shown in Figs. 5(a), this pre-strained sample has a larger coercivity than the ordinary one at the as-deposited state. It may be caused by a joint effect of the phase-transition process from R to M1 causing an enhancement of PMA and the higher grown temperature resulting in a better film quality. With the temperature increase, the hysteresis loop of the pre-strained sample change with a same trend of the ordinary one, but the effect is obviously different. The variation of Hs has been summarized in Fig.5 (b) as a function of temperature. For the ordinary sample, $H_S$ has increased about 660Oe up to 353 K, while the value is only about 170 Oe for the pre-strained one. The inset gives the difference ratio between these two sample, calculated as η=($H_{S, RT}$-$H_{S\ 373K}$)/ $H_{S\ 373K}$, which value reaches as high as 620%. The temperature range where significant difference can be observed between these two samples is above 338 K, i.e. VO$_2$ is at R state. It reflects that if the interfacial strain effect is pre-release, the modulation effect would be highly weakened. Considering the saturation field takes an important role in magnetic switching, this huge tunable feature has important research value in the field of magnetic recording.

To make a conclusion, we have investigated the reversible magnetic anisotropy modulation effect in the heterostructure of $(Co/Pt)_2/VO_2$ system, where a strongly correlated electron system $VO_2$ is used to applied an interfacial strain by its phase-transition. The energy density change of perpendicular anisotropy and interfacial anisotropy reaches 38 kJ/m$^3$ and 100 mJ/m$^2$, respectively. We have also shown that a strain buffered insertion layer and pre-strained process could influence the modulation effect, suggesting that the interfacial strain between $VO_2$ and PMA layers is responsible for the observed modulation. The present result is expected to provide new guidelines for the improvement of the efficiency of PMA modulation and benefits the development of spintronic research.

**Acknowledgement:**

This work was supported by the National Natural Science Foundation of China (Nos. 51602013, 11804016, 61704005, 61571023, and 61627813), the International Collaboration 111 Project (No. B16001), the Beijing Natural Science Foundation (No. 4162039), the China Postdoctoral Science Foundation (No. 2018M631296) and the Beijing Advanced Innovation Center for Big Data and Brain Computing (BDBC).

**Figures and Legends**

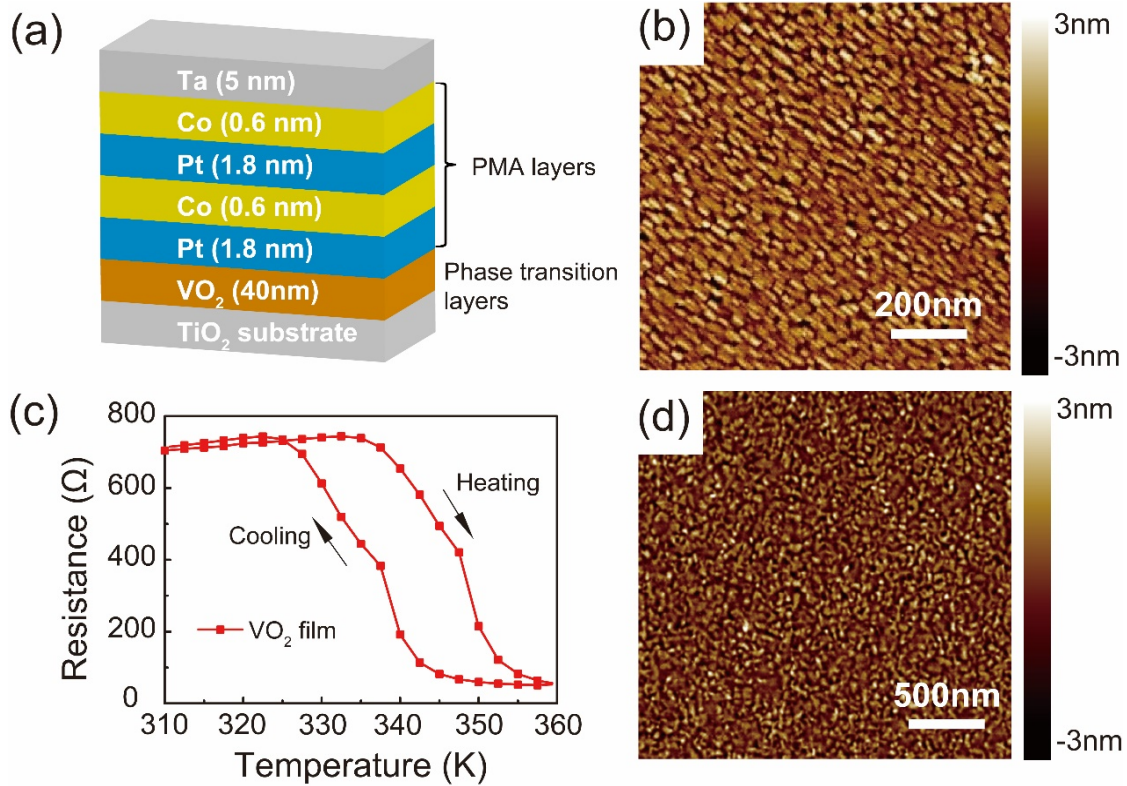

FIG.1. (a) Schematic drawing of the PCMA heterostructure. (b) AFM image of the heterostructure film after VO$_2$ deposition. (c) Temperature dependence of VO$_2$ resistance used in this work. (d) AFM image of the film after all layers are deposited.

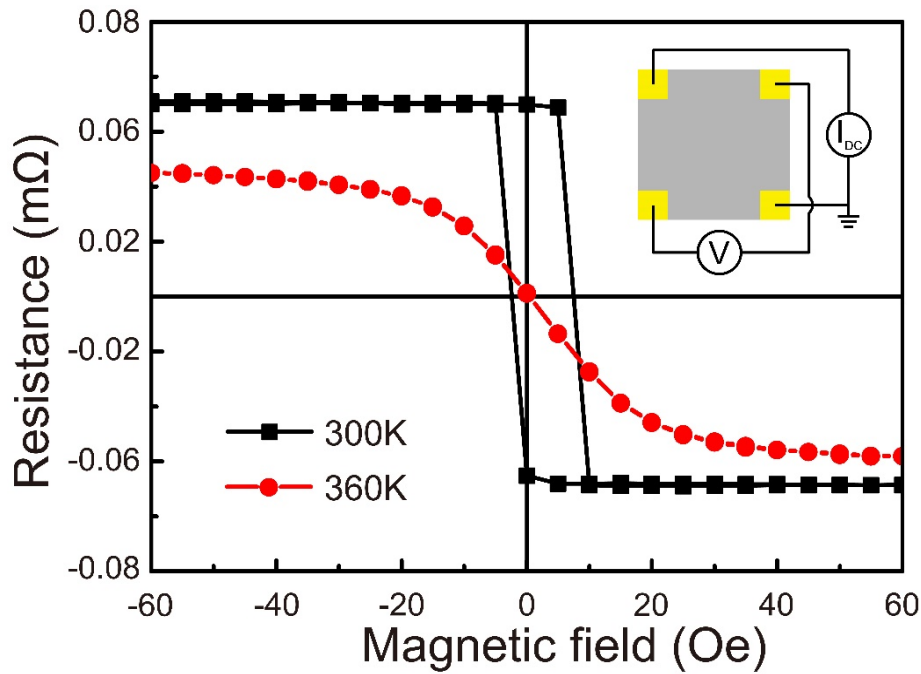

FIG. 2. The magnetic field dependence of the Anomalous Hall resistivity measured at 300 K and 360K.

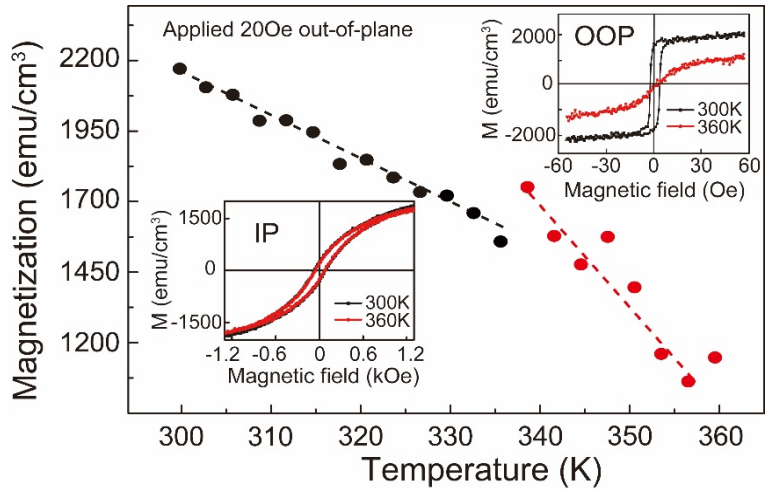

FIG. 3. The temperature dependence of the magnetization measured with 2Oe magnetic field applied out-of-plane. The inset gives the hysteresis loops IP and OOP measured at 300 and 360 K.

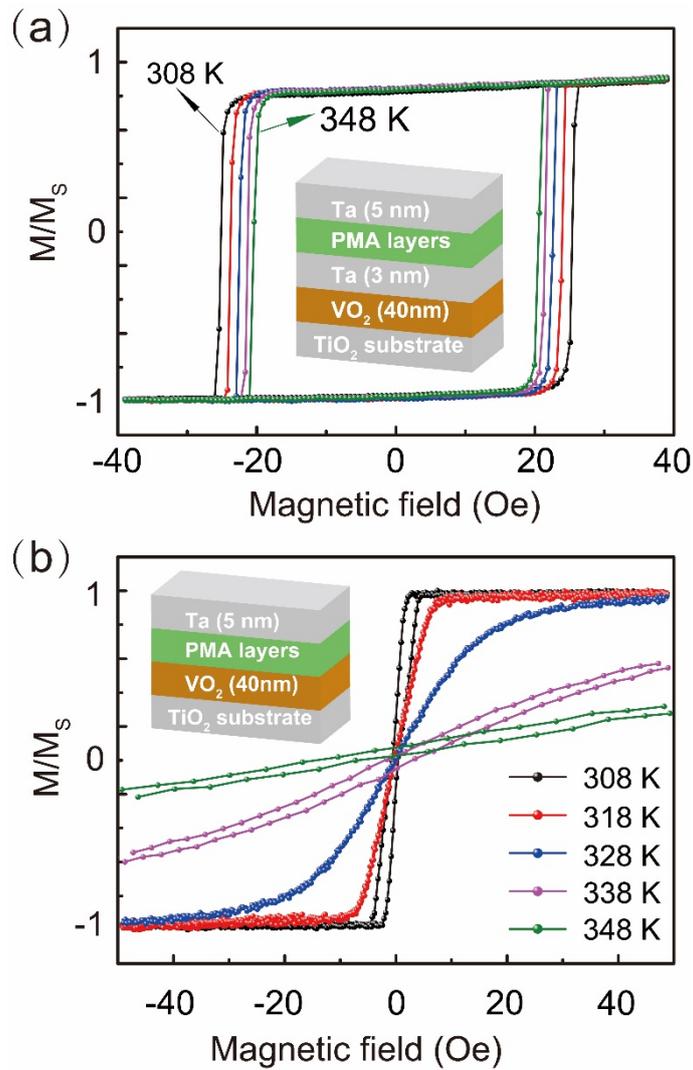

FIG. 4. Out-of-plane hysteresis loops measured by MOKE under different temperature. (a) Strain-buffered sample with 3nm Ta inserted between $VO_2$ and PMA layers. (b) Ordinary sample with the PMA layers deposited at room temperature.

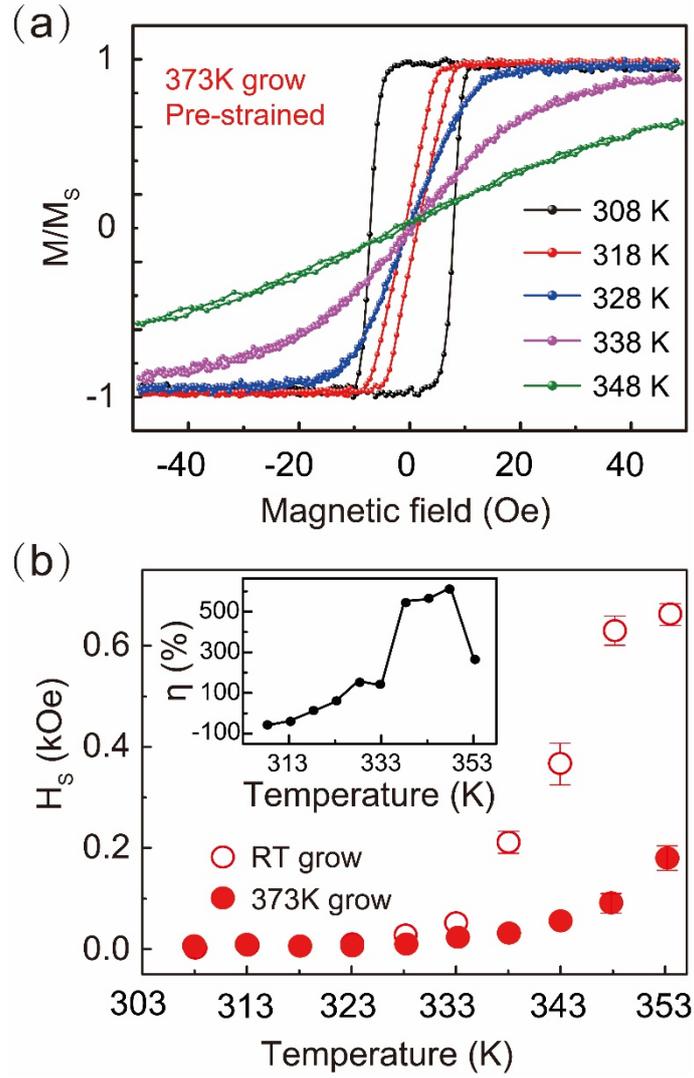

FIG. 5. (a) Pre-strained sample with the PMA layers deposited at 373K. (b) $H_S$ change comparison between the ordinary and pre-strained samples. The inset gives η variation as a function of temperature.

TABLE 1. Magnetic properties of $TiO_2$ (100)/$VO_2$ (40 nm)/[Pt (1.8nm)/Co (0.6 nm)]$_2$/Ta (5nm)

| Temperature | $M_S$ | $K_{eff}$ | $K_{eff} t_{Co}$ | $K_i$ |
| --- | --- | --- | --- | --- |
| K | emu/cm$^3$ | erg/cm$^3$ | erg/cm$^2$ | erg/cm$^2$ |
| 300 | 2030 | 7.6×10$^5$ | 0.09 | 3.1 |
| 360 | 2000 | 3.8×10$^5$ | 0.05 | 3.0 |